\documentclass[twocolumn,showpacs,preprintnumbers,amsmath,amssymb,prl]{revtex4}

\usepackage{graphicx}
\usepackage{dcolumn}
\usepackage{bm}

\begin{document}
\newcommand{\rar}{$\rightarrow$}
\newcommand{\lrar}{$\leftrightarrow$}

\newcommand{\beq}{\begin{equation}}
\newcommand{\eeq}{\end{equation}}
\newcommand{\bea}{\begin{eqnarray}}
\newcommand{\eea}{\end{eqnarray}}
\newcommand{\Req}[1]{Eq.\ (\ref{E#1})}
\newcommand{\req}[1]{(\ref{E#1})}
\newcommand{\degree}{$^{\rm\circ} $}
\newcommand{\pcite}{\protect\cite}
\newcommand{\pref}{\protect\ref}
\newcommand{\Rfg}[1]{Fig.\ \ref{F#1}}
\newcommand{\rfg}[1]{\ref{F#1}}
\newcommand{\Rtb}[1]{Table \ref{T#1}}
\newcommand{\rtb}[1]{\ref{T#1}}

\title{The Worm-Like Chain Theory And Bending Of Short DNA}

\author{Alexey K. Mazur}
\email{alexey@ibpc.fr}
\affiliation{CNRS UPR9080, Institut de Biologie Physico-Chimique,
13, rue Pierre et Marie Curie, Paris,75005, France.}


\begin{abstract}

The probability distributions for bending angles in double helical DNA
obtained in all-atom molecular dynamics simulations are compared with
theoretical predictions. For double helices of one helical turn
and longer the computed distributions remarkably agree with the
worm-like chain theory and qualitatively differ from predictions of
the semi-elastic chain model. The computed data exhibit only small
anomalies in the apparent flexibility of short DNA and cannot account
for the recently reported AFM data (Wiggins et al, Nature
nanotechnology 1, 137 (2006)). It is possible that the current
atomistic DNA models miss some essential mechanisms of DNA bending on
intermediate length scales.  Analysis of bent DNA structures reveals,
however, that the bending motion is structurally heterogeneous and
directionally anisotropic on the intermediate length scales where the
experimental anomalies were detected. These effects are essential for
interpretation of the experimental data and they also can be
responsible for the apparent discrepancy.

\end{abstract}


\maketitle

The bending dynamics of double helical DNA on length scales of several
helical turns plays a key role in many cellular processes
\cite{Widom:01}. This length scale is also crucial for the coupling
between the atomistic DNA structure and its macroscopic mechanics.
Long DNA double helices are well described by the classical worm-like
chain (WLC) model with a persistence length of $A\approx$50 nm
\cite{Cantor:80}, but for chain lengths shorter than $A$ the validity
of the WLC theory is uncertain \cite{Cloutier:05,Du:05a}. Anomalously
high flexibility was sometimes observed for double helices as short as
40 nm that exhibited experimental cyclization rates several orders of
magnitude beyond the WLC predictions \cite{Cloutier:05}. Until now, a
few theories proposed to account for these anomalies did not converge
to a consensus interpretations \cite{Yan:04,Wiggins:05,Wiggins:06b}.
Very recently, AFM imaging experiments with direct counting of DNA
bends in planar deposits revealed strong deviations of bend angle
distributions from the WLC theory for chain length 5-20 nm
\cite{Wiggins:06a}. In addition, these data disagreed with all
alternative theories except the so-called semi-elastic chain model
(SEC) \cite{Wiggins:06b}.  However, statistical analysis of molecular
dynamics (MD) trajectories indicates that atomistic DNA models agree
with the WLC theory already for double helices of 1-2 helical turns
\cite{Mzbj:06}. This apparent controversy suggests either that earlier
analysis of MD data was not complete or that MD simulations miss some
essential molecular mechanisms of DNA bending.

In this Letter we present the first accurate comparison of analytical
theories with DNA bend angle distributions observed in realistic MD
simulations and try to get insight in the possible origin of anomalies
in the apparent DNA flexibility on intermediate length scales. The MD
data used here were obtained in long MD simulations of double helical
DNA fragments of 25 base pairs (bp) with the AT-alternating sequence.
For our present purposes this model system can be considered as
homopolymer. The same data were already used for evaluation of elastic
parameters of atomistic DNA models \cite{Mzbj:06} and we refer the
reader to this earlier paper for simulation details. Three MD
trajectories denoted AT25a-c, respectively, differed by hydration
conditions as well as the number of degrees of freedom in the DNA
duplexes. In AT25a (16 ns) DNA was modelled with all degrees of
freedom in a rectangular water box with a neutralizing number of
sodium ions. In AT25b (28 ns) the hydration conditions were the same
as in AT25a, but the duplex was modeled with fixed geometry of
chemical groups, rigid bases and only partially flexible backbone.  In
AT25c (120 ns) the minimal B-DNA model was used, with semi-implicit
treatment of solvent as described earlier \cite{Mzjacs:98}. The
analysis below uses the data from trajectory AT25a by default, but the
results and conclusions were qualitatively similar for all three
trajectories. For better sampling, statistical analysis included all
internal fragments of a given length in the 25-mer DNA.

\begin{figure}[htb]
\includegraphics [width=8.6cm]{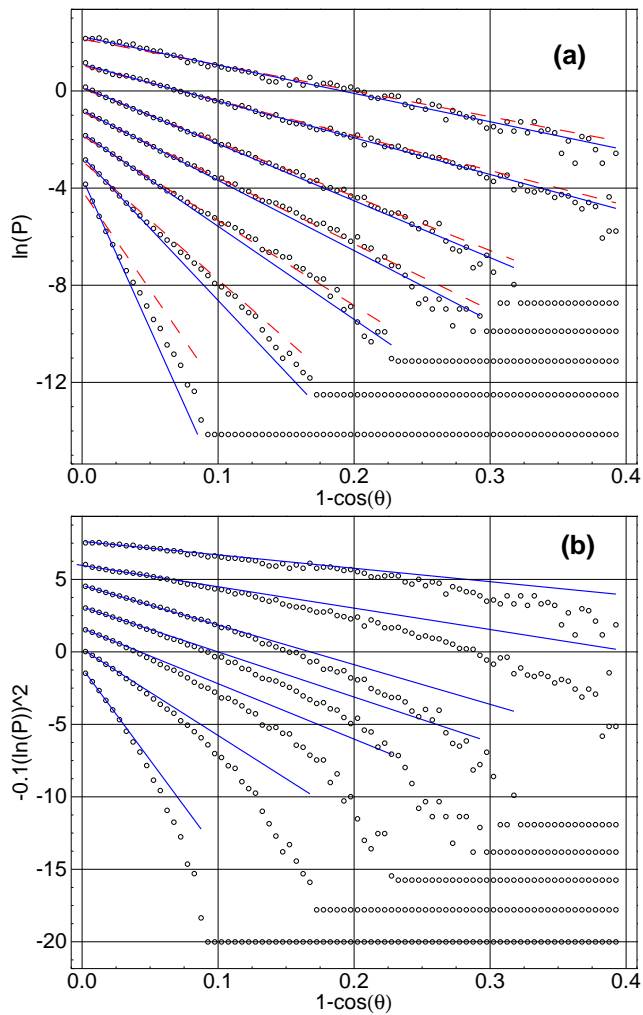}
\caption{\label{Ff1} Color online. Probability distributions of bend
angles in DNA fragments of $N$=4, 6, 8, 10, 12, 18, and 24 bp.
Data counts were accumulated in 80 windows evenly spaced on
$0.6<\cos\theta<1$. (a) The y-coordinate is chosen to linearize the
WLC angle distributions.  Red dashed lines correspond to WLC
distributions with $A$=80 nm (the value earlier obtained from the same
data by other methods \cite{Mzbj:06}). Blue solid lines represent
linear regression analysis of the initial decrease down to $e^{-2}$
including at least 5 points. All plots are consecutively shifted by
one for clarity.  (b) The y-coordinate is chosen to linearize the SEC
angle distribution.  Blue solid lines represent linear regression
analysis of the initial decrease down to $e^{-3}$ including at least 5
points. All plots are consecutively shifted by 1.5.} \end{figure}

The WLC theory approximates the bending probability by
the Gaussian distribution
\cite{Cantor:80,Landau:76}
\beq\label{Ew0}
  w_0\sim\exp(-\frac{A}{2L}\theta^2)
\eeq
where $\theta$ is the bend angle, $L$ is the length of the DNA
fragment, and $A$ is the persistence length. Equation \req{w0} is valid
for sufficiently small $L$ and $\theta$, which is very well fulfilled
for practical MD simulations. Assuming spherical isotropy of vector
orientations randomly sampled in 3D space the differential of the
probability distribution for small bend angles can be written as
\bea\label{EP}
 dP\sim\exp(-\frac{A}{2L}\theta^2)\sin\theta d\theta\approx\nonumber\\
     \exp[-\frac{A}{L}(1-\cos\theta)]d(1-\cos\theta).
\eea
Equation \req{P} indicates that the plots of the WLC probability
density would give descending straight lines in coordinates: $(\ln P)\
vs\ (1-\cos\theta)$, with variation of slopes governed by the DNA
persistence length.

The SEC theory \cite{Wiggins:06b} does not predict the shapes of angle
distributions for all chain lengths. It postulates that there exists
an intermediate DNA length $l_0$ for which the bending probability is
exponential
\beq\label{Ew1}
  w_1\sim\exp(-\alpha\theta)
\eeq
Parameters $l_0$ and $\alpha$ have been estimated as 2.5 nm and 6.8,
respectively \cite{Wiggins:06a}. For $L<l_0$, the SEC behavior is undefined,
whereas for $L>l_0$ all its statistical properties converge to the WLC model
due to the central limiting theorem. By applying the considerations of
\Req{w0}-\req{P} we see that the SEC probability density would give a
descending straight line for $L=l_0$ in coordinates: $-(\ln P)^2\ vs\
(1-\cos\theta)$.

Figure \rfg{f1} compares the shapes of DNA bend angle distributions
obtained in the most detailed representation (AT25a) with the WLC and
SEC theories. The agreement with the WLC theory is remarkable. The MD
distributions in \Rfg{f1}a are nearly linear and for $N\geq 6$ they
agree very well with the WLC theoretical predictions shown by the
dashed red straight lines. All these theoretical traces correspond to
one and the same persistence length equal to the earlier reported
value obtained from these MD data by different methods \cite{Mzbj:06}.
In contrast, \Rfg{f1}b shows that these MD distributions
systematically deviate from the SEC theoretical predictions.
Qualitatively, the deviations in \Rfg{f1}b are similar for all chain
length suggesting that the SEC critical length $l_0$ can only be
smaller. At the same time, \Rfg{f1}a reveals for DNA fragments of 4-10
bp noticeable upward deviations of the tails of the distributions with
respect to the linear regression lines obtained for small angles only,
which qualitatively agrees with the SEC assumption. Similar patterns
were obtained for trajectories AT25b and c. The deviations from the
WLC theory revealed in \Rfg{f1}a are smaller than necessary for the
correspondence with the SEC model, nevertheless, they are reproducible
and require explanation.

\begin{figure}[htb]
\includegraphics [width=8.6cm]{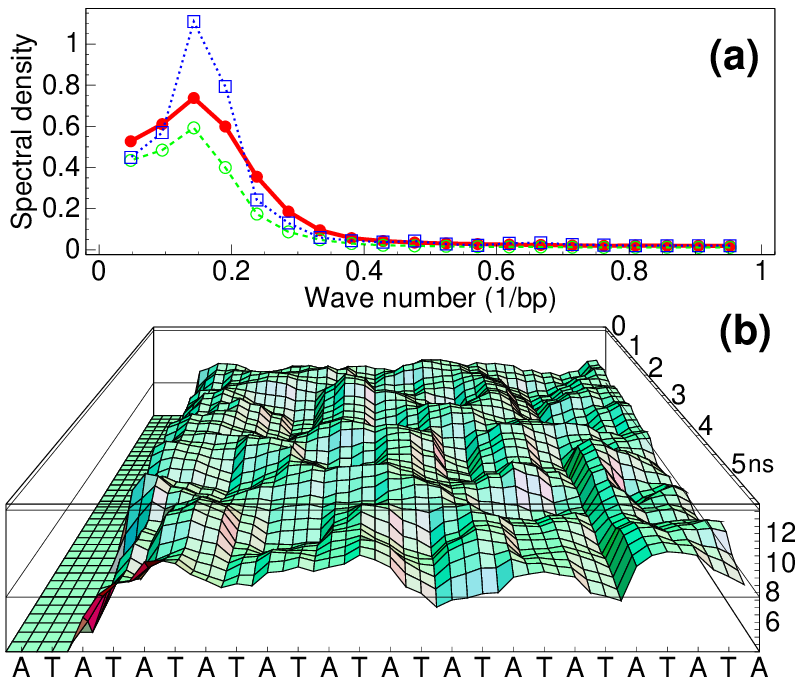}
\caption{\label{Ff2} Color online. (a) The average Fourier spectra of the
profiles of the minor groove for three trajectories of the 25-mer
AT-alternating fragment. AT25a - solid red line and filled circles.
AT25b - dashed green line and open circles; AT25c - dotted blue line
and open squares.  The width of the minor groove was measured as
described elsewhere \cite{Mzjmb:99} for all saved states and used for
spectral analysis.  Thus obtained spectral densities were averaged
over the entire trajectories. (b) An example of time evolution of the
minor groove profile. The surface plot is formed by 40 evenly spaced
profiles taken on a sample 6 ns interval from trajectory AT25a. Every
profile was averaged over 500 ps.} \end{figure}

The non-linearities seen in \Rfg{f1}a can be rationalized if the
population of DNA fragments of a given length is heterogeneous as
regards elastic properties. For instance, if 50\% of 10-mers are rigid
while the rest 50\% are more flexible the probability density for
small angles would decrease more rapidly than expected for the average
persistence length (see solid straight lines in \Rfg{f1}a) In
contrast, large bends can still occur due to the second group and the
probability density plots in \Rfg{f1}a would deviate upward. For
homopolymer DNA such heterogeneity can originate from the intrinsic
frustration in the B-DNA structure as suggested by the compressed
backbone hypothesis (CBH) \cite{Mzjacs:00,Mzpre:02,Mzbchem:04}, and \Rfg{f2}
reveals its possible mechanism. It is seen that the minor groove of
this DNA fragment exhibits quasi-regular modulations with a
distinguished period of about 8 bp. This behavior originates from the
compressed state of the backbone in the B-DNA structure as described
elsewhere \cite{Mzjacs:00,Mzpre:02}. The DNA bending and groove
dynamics are not directly linked, but uneven minor groove profile can
facilitate or hamper bending in certain directions. Now consider the
ensemble of DNA fragments of 10 bp, for instance.  This length is
sufficient to accommodate one widening of the minor groove flanked by
two narrowings, or one narrowing flanked by two widenings.  If the
bendability in these two sub-ensembles strongly differ one would
expect to observe the kind of deviation seen in \Rfg{f1}a for 10-mers.
More precisely, \Rfg{f2} indicates that a DNA fragment of length $L$
has at least $n=8$ equally probable, but qualitatively different
configurations of the minor groove profile.  Assuming that in each of
these sub-populations bending is approximately Gaussian the overall
bending probability can be expanded as

\beq\label{EwL}
  w(L)\sim\sum_{i=1}^{n}\exp[-\alpha_i(L)(\theta-\theta_i)^2].
\eeq
The possibility of static bends, that is $\theta_i\neq 0$, should not
be excluded here. When $L$ is comparable with the period of groove
modulations the values of $\alpha_i$ and $\theta_i$ can vary giving a
non-Gaussian overall distribution of bend angles. For longer chains the
heterogeneity of $\alpha_i$ and $\theta_i$ should gradually disappear.

\begin{figure}[htb]
\includegraphics [width=8.6cm]{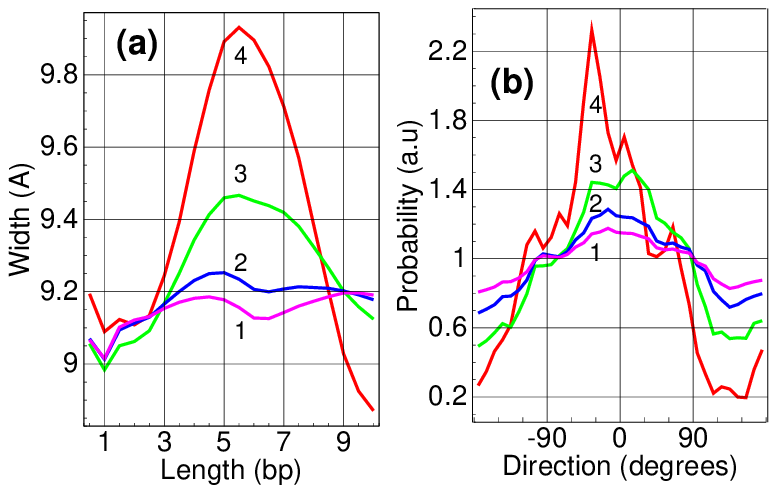}
\caption{\label{Ff3} Color online. (a) The average minor groove
profiles evaluated in 10 bp DNA stretches sorted by the amplitude of
bending.  1 (magenta) - overall ensemble average; 2 (blue) - fragments
bent by more than 10$^\circ$; 3 (green) - fragments bent by more than
20$^\circ$; 4 (red) - fragments bent by more than 30$^\circ$. The
traces were smoothed with a sliding window of 2 bp to remove the
sequence effect. (b) Radial distributions of bending direction in
sub-ensembles of 10 bp fragments used in plate (a). The bending
direction is defined by the polar angle of the projection of the
opposite ends upon the middle plane as described elsewhere
\cite{Mzjacs:00}. The zero angle is chosen to corresponds to a planar
bend towards the major groove in the middle of the 10-mer.  Colors and
numbering are same as in plate (a). All plots are normalized by the
small radial distribution corresponding to bend angles below
10$^\circ$.  } \end{figure}

To verify the foregoing interpretation we compared the minor groove
profiles in weakly and strongly bent conformers. The results are shown
in \Rfg{f3}a. On average, the minor groove of 10-mer fragments is
relatively even, which might be expected from the dynamics shown in
\Rfg{f2}b. In contrast, strongly bent conformers tend to have a
widening in the middle flanked by narrowings. This formally proves
that parameters $\alpha_i$ and $\theta_i$ in \Req{wL} vary, which
explains deviations from linearity seen in \Rfg{f1}a. In addition,
\Rfg{f3}b shows that strong bends in 10-mers preferably occur towards
the the major groove in the middle, that is away from the central
widening and towards the flanking narrowings of the minor groove in
\Rfg{f3}a. Given the pattern displayed in \Rfg{f2}b and \rfg{f3}a,
this effect does not seem surprising since it agrees with the earlier
known general trends characteristic of curved DNA \cite{Crothers:99}.
However, it points to another possible source of anomalies in the
experimental distribution of bend angles in planar AFM images
\cite{Wiggins:06a}.  The 10-mer fragments of a long double helix
deposited on a plane face it by all their sides with equal
probability. If the pattern revealed in \Rfg{f3} remains valid for
planar depositions, 10-mer fragments with the center of the major
groove turned towards or away from the plane should on average look
stiffer than those where the preferred bend direction revealed in
\Rfg{f3}b is parallel to the plane. Indeed, bends in directions
perpendicular to the plane can be completely blocked or hindered, but
anyway they are not detectable in AFM images. This effect can be
expected to persist for DNA length of 10-20 bp and it can introduce an
additional hidden heterogeneity in the ensembles of planar DNA
segments of a given length.

The character of bending revealed in \Rfg{f2} and \rfg{f3} indicates
also that the specific atomistic structure of bent DNA may interfere
with in the very process of DNA deposition on a plane. Indeed, the
preferred period of 8 bp for the minor groove modulations revealed in
\Rfg{f2}a together with the pattern shown in \Rfg{f3} indicates that
strong bends in the neighboring stretches of long DNA tend to occur in
nearly perpendicular directions. This short-range correlation is
incompatible with planar geometry and it should be somehow cancelled
during deposition and equilibration of 3D DNA on a plane, which can
produce unpredictable effects upon the distribution of bend angles.

The results presented in this Letter demonstrate that the probability
distributions for bending angles in short stretches of double helical
DNA obtained with the most accurate currently used atomistic models
remarkably agree with the WLC theory already for lengths of about one
helical turn. Recent molecular dynamics (MD) simulations of circular
DNA indicated that strong bends in double helices may occur due to
localized kinking, rather than smooth curvature corresponding to the
WLC model \cite{Lankas:06,Ruscio:06}. In both cases, however, the
kinks could be due to external factors like excessive bending strain
\cite{Lankas:06} or protein-DNA contacts \cite{Ruscio:06}.  In the
present studies, the contribution kinks was not significant even in
the strongest bends. It is hardly possible that longer MD simulations
would change this conclusion for the range of bend angles sampled
here. The ensemble of earlier MD studies of free DNA
\cite{Cheatham:00} evidences that breaking of base pair stacks like in
Ref. \onlinecite{Lankas:06} is never observed probably because of high
energy barriers. Therefore, to affect the populations of bend angles in
the range sampled here, such states have to be very long-living, and
the question arises why such states were not be detected in earlier
NMR and X-ray studies \cite{Dickerson:99}. At the same time, our
results do not rule out the possibility of kinked states for very large
angles not sampled here.

Our data do not agree with the recent report on anomalously high
flexibility of DNA fragments of 5-17.5 nm detected by AFM experiments,
as well as the SEC theory that explained these experimental data. It
is possible that the present day atomistic DNA models do not reproduce
some essential aspects of DNA bending on intermediate length scales.
Analysis of bent DNA structures reveals, however, that, in spite of
the good agreement with the WLC theory, the bending motion is
structurally heterogeneous and directionally anisotropic on the
intermediate length scales where the experimental anomalies were
detected. These effects are essential for interpretation of the
experimental data and they also can be responsible for the apparent
discrepancy.

\end{document}